# Constellations of Colour: Mapping Semantic Neighbourhoods in CLIP

**Matt DesLauriers**
University of the Arts London
Creative Computing Institute
London, UK
m.deslauriers0620251@arts.ac.uk

**ABSTRACT**
How do vision-language models organise the cultural and perceptual dimensions of colour language? This pictorial presents a research-through-design visualisation of colour semantics in CLIP, a pretrained vision-language model. A dataset of 2,500 named colours is embedded with CLIP's text encoder, projected to a 2D map, and connected via a thresholded minimum spanning tree to create disjoint "constellations" of related colour terms.

The resulting maps offer a model for navigating the semantic and perceptual relationships between colour terms. The visualisations reveal gradients along familiar axes (hue, lightness, chroma), as well as semantic neighbourhoods, including botanical language (hyacinth, cornflower, tulip), fictional references ("Robin Hood green", "warlock purple"), and food (butterscotch, toast, toffee). These visualisations give designers and creatives a map for exploring colour linguistically and building palettes by semantic association, pointing towards interpretable, artist-facing colour tools derived from foundation models.

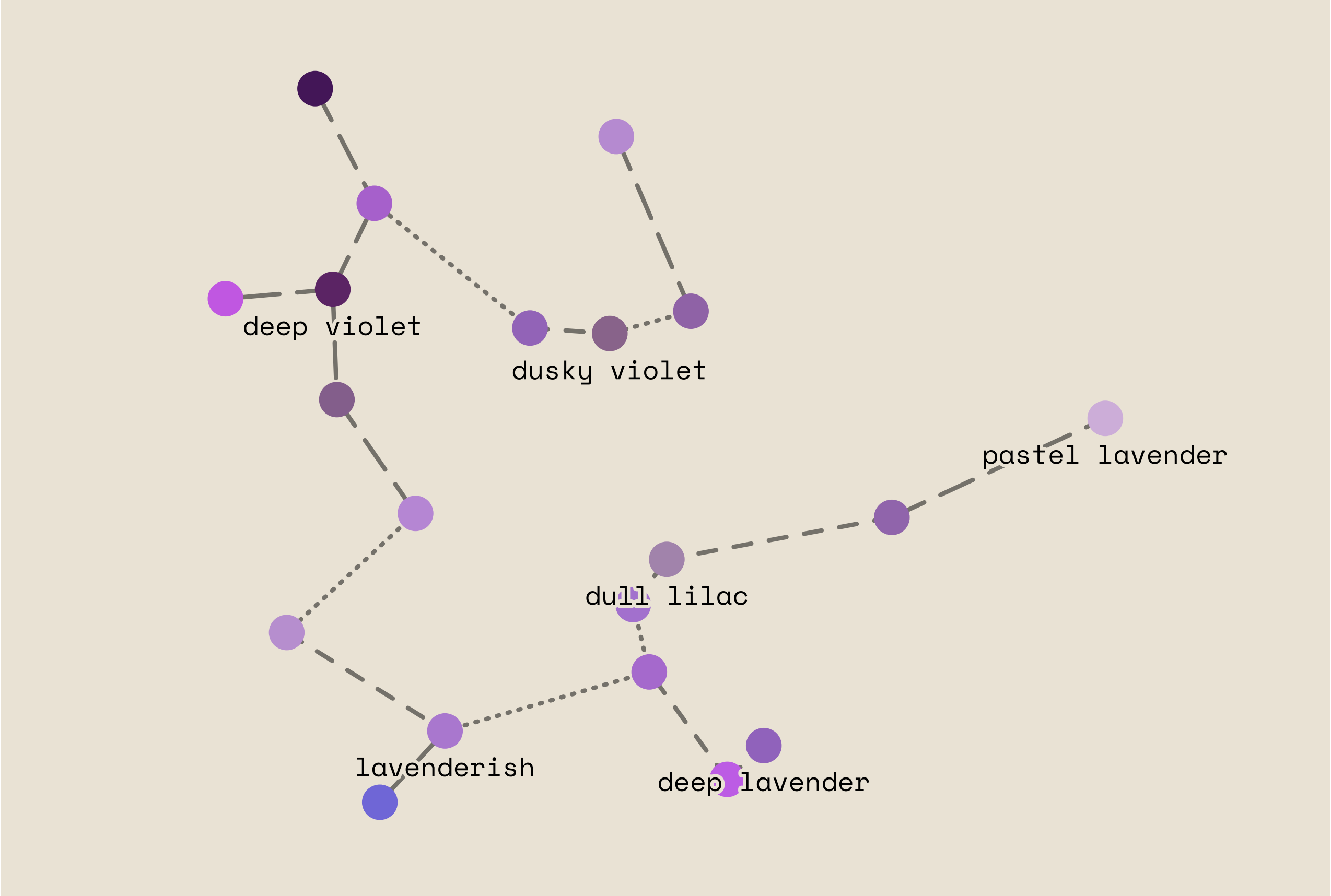

↑ A constellation of lavender- and violet-related colour terms, mapped by cosine similarity in CLIP embedding space.

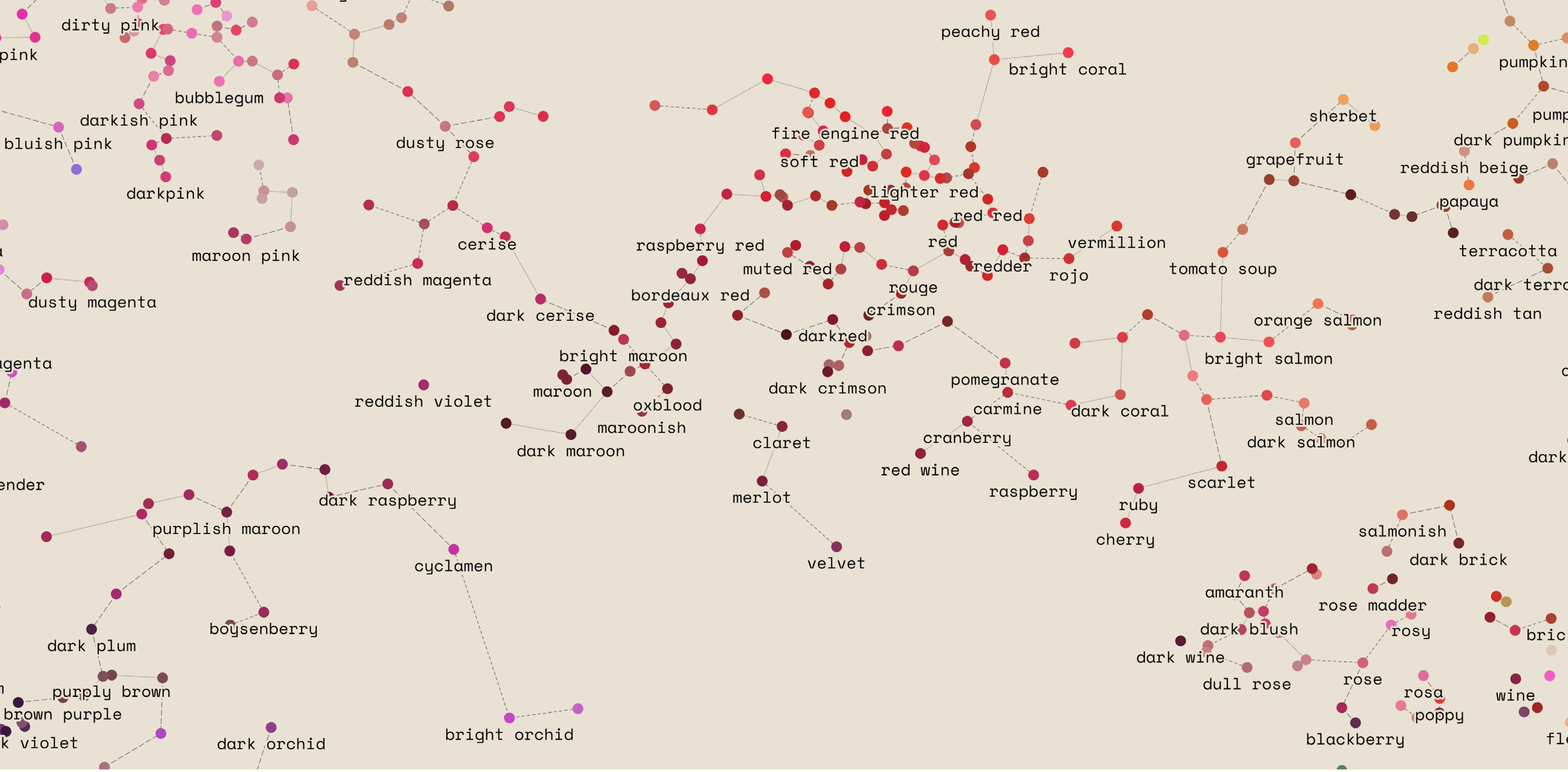

**MAPPING COLOUR IN CLIP SPACE**
CLIP is a pretrained vision-language model that learns a joint embedding space aligning text and images, encoding both perceptual and linguistic associations into high-dimensional vector embeddings [1]. To examine how colour is organised in CLIP's 512D embedding space, we curate a dataset of 2,500 popular colour terms from the 2010 XKCD online colour naming survey [2]. The terms are embedded with CLIP's text encoder and projected into 2D using UMAP [3], producing a map that approximately preserves local similarity in embedding space.

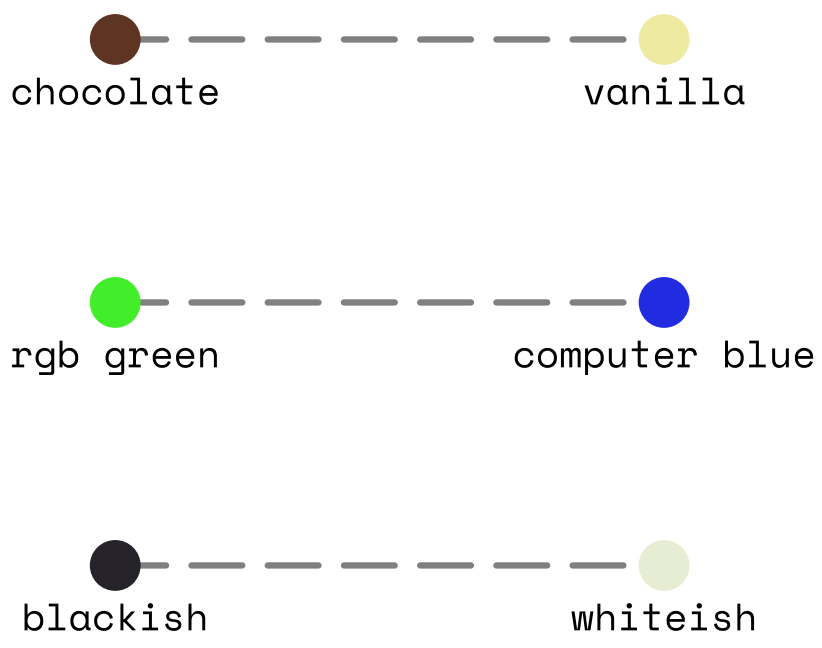

↑ We use colour naming and data visualisation as an artistic probe into CLIP's internal representation of language and perception. Shown above is a close-up crop of the larger map, offering a practice-based, interactive, and visual approach to model interpretation and explainability.

← Our model can also reveal interesting relationships between colour terms. Shown here are pairs of colour terms that are close in CLIP's embedding space, and therefore semantically related, but distant by a perceptual colour difference metric.

## DATA COLLECTION AND METHODOLOGY

The online XKCD colour survey presented viewers with random RGB colours and asked them to label each. Over 3 million responses were collected. Names varied from common terms such as “blue” and “red,” to more obscure labels like “butterscotch,” “shamrock,” and “brorange” (a portmanteau of brown-orange). Although colour categorisation varies considerably across languages and cultures [6], participants in this survey were largely English-speaking and internet-literate, and the final vocabulary carries this bias.

After handling typos, near-duplicates, outliers, spam, and profanities, unique labels are sorted by vote count to produce a set of 2,500 popular colour terms. We use OKLab [4], a perceptually uniform colour space widely supported in web browsers, to compute a mean display colour for each term.

After embedding with CLIP’s text encoder and projecting to 2D with UMAP, we construct constellation plots to navigate the space by neighbourhoods of related terms. A minimum spanning tree connects colour terms by their cosine similarity in CLIP embedding space. Edges above a distance threshold are cut, forming a collection of disjoint graphs. These often form distinct regions organised by hue, lightness, chroma, and shared semantic reference. Edges are rendered with dash patterns based on Euclidean distance in OKLab space, providing a visual hint for perceptual similarity between nodes: short dashes represent colours close in perceptual space, long dashes represent colours far apart.

Once fitted, the UMAP projection can transform additional CLIP embeddings onto the same 2D space. Unseen colour terms, and even non-colour terms, can be mapped to the existing layout, where they often fall in semantically and perceptually related regions.

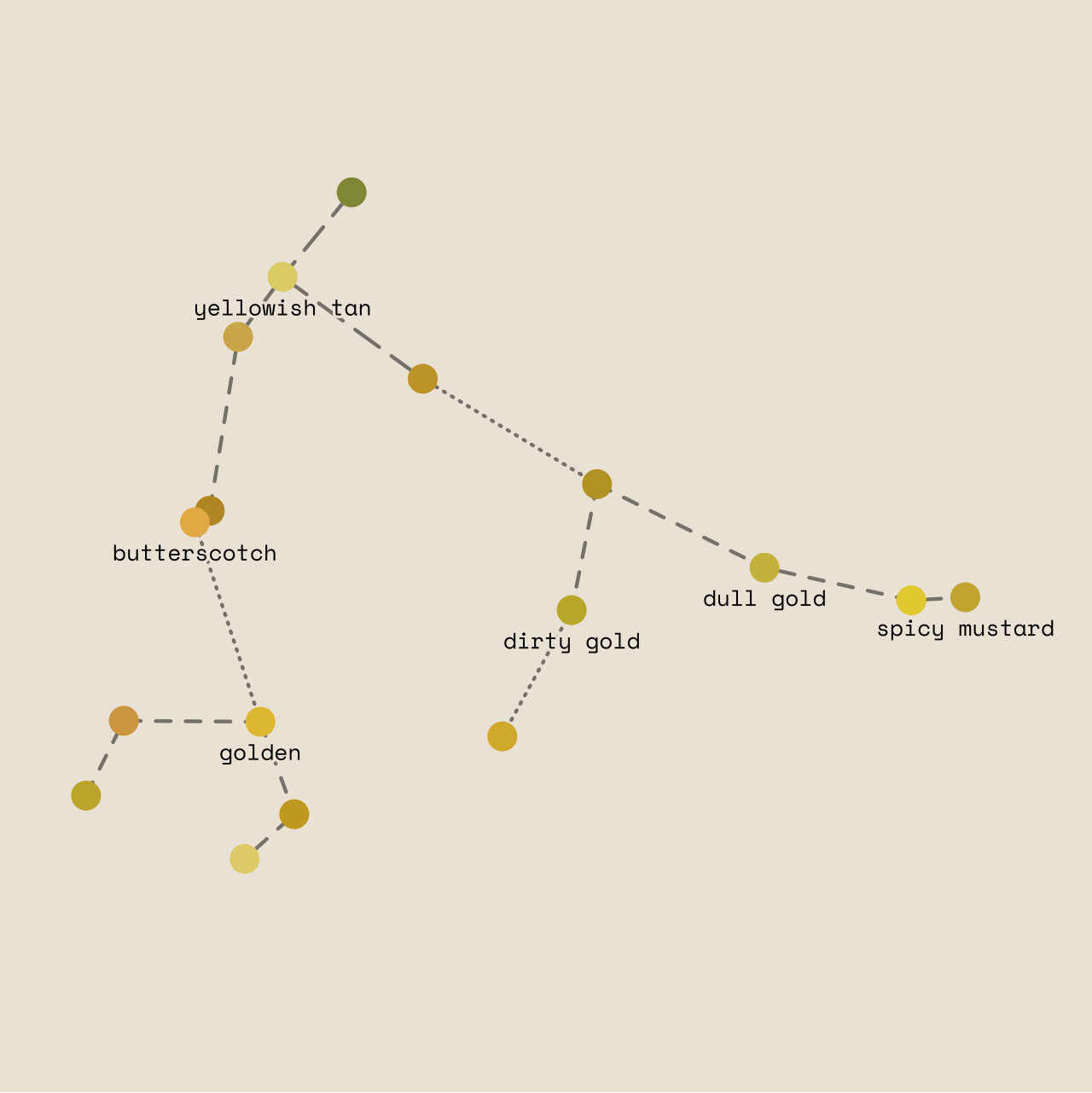

↗ A constellation plot of a yellow-gold neighbourhood, linking culinary terms (“butterscotch,” “spicy mustard”) with abstract descriptors (“golden,” “dull gold”).

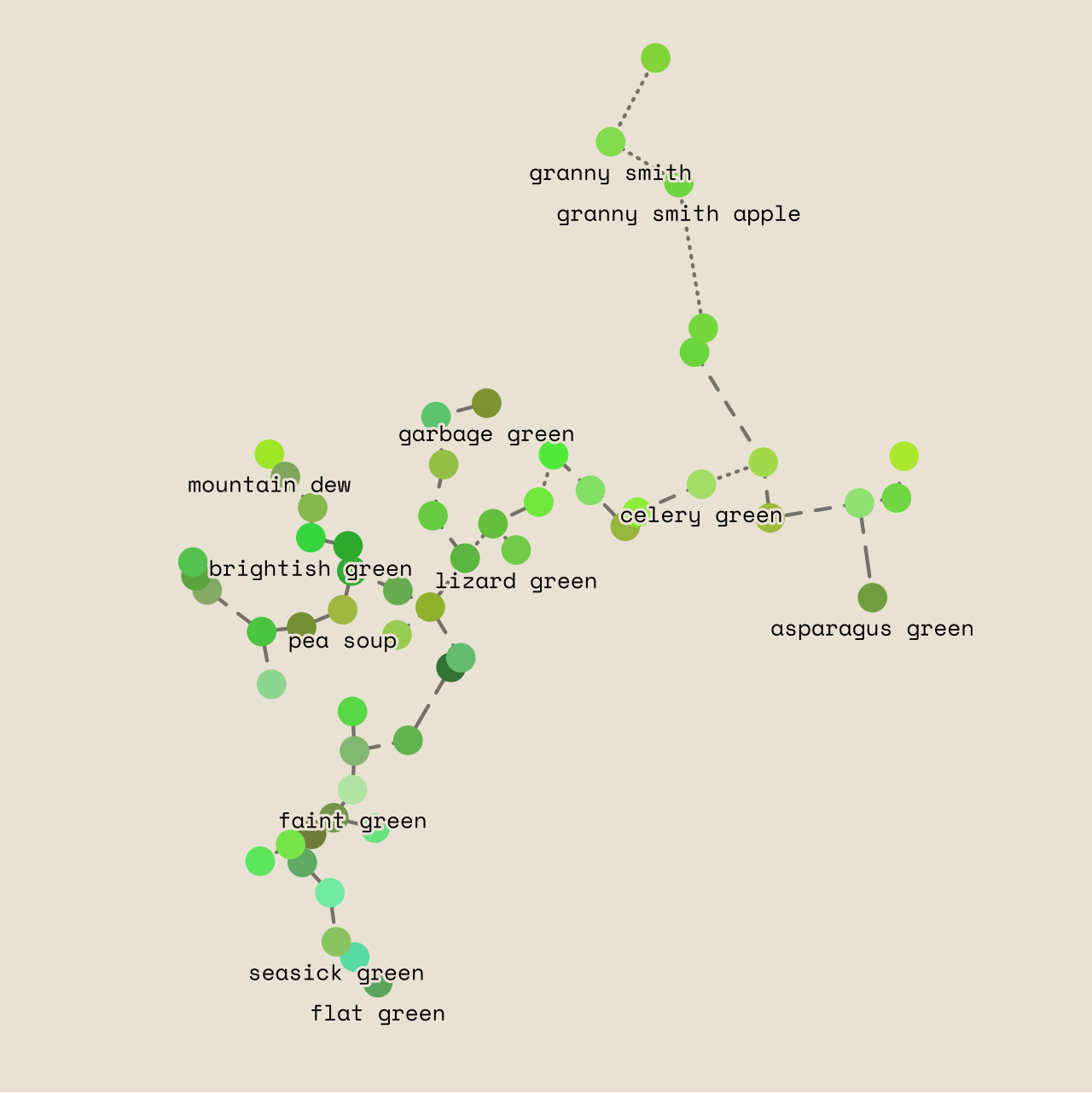

↗ A constellation plot of greens, loosely connecting the brand name label “mountain dew” with green foods (“pea soup”, “granny smith,” “asparagus green”).

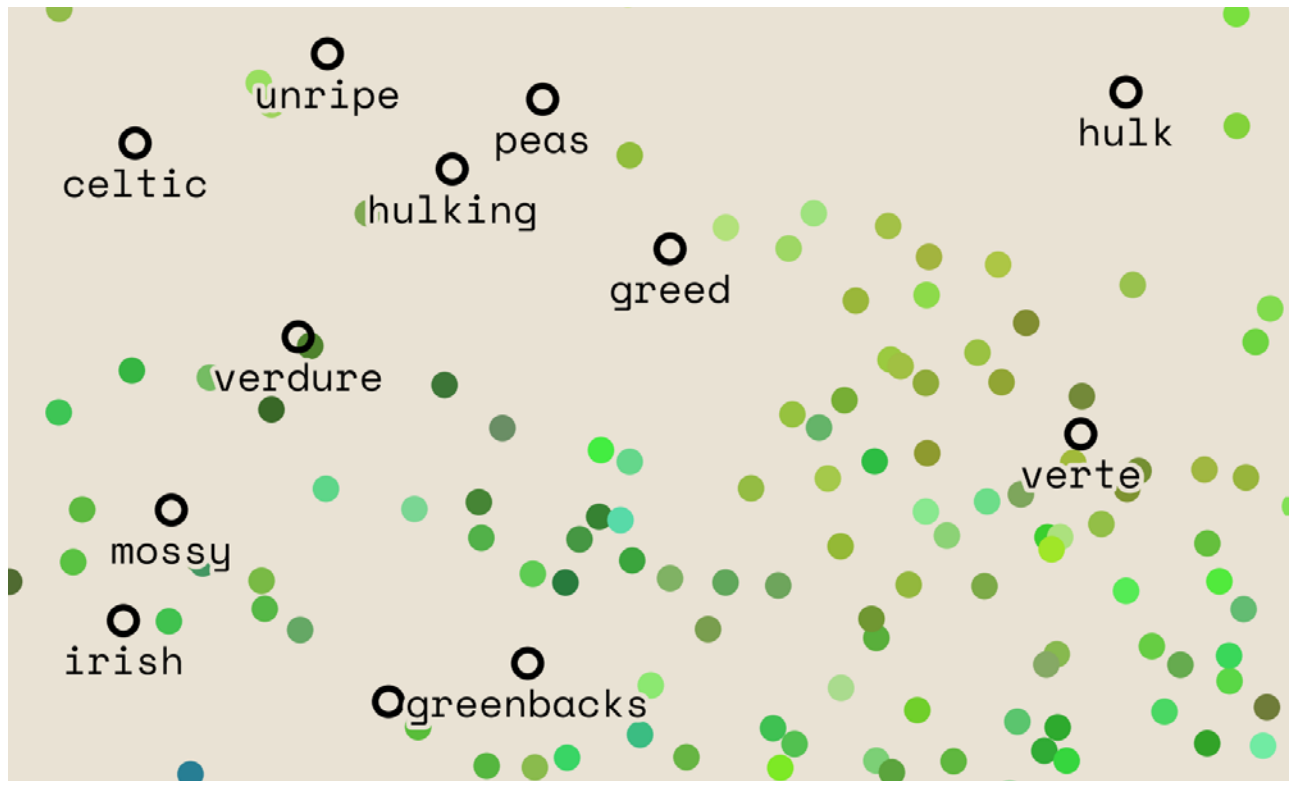

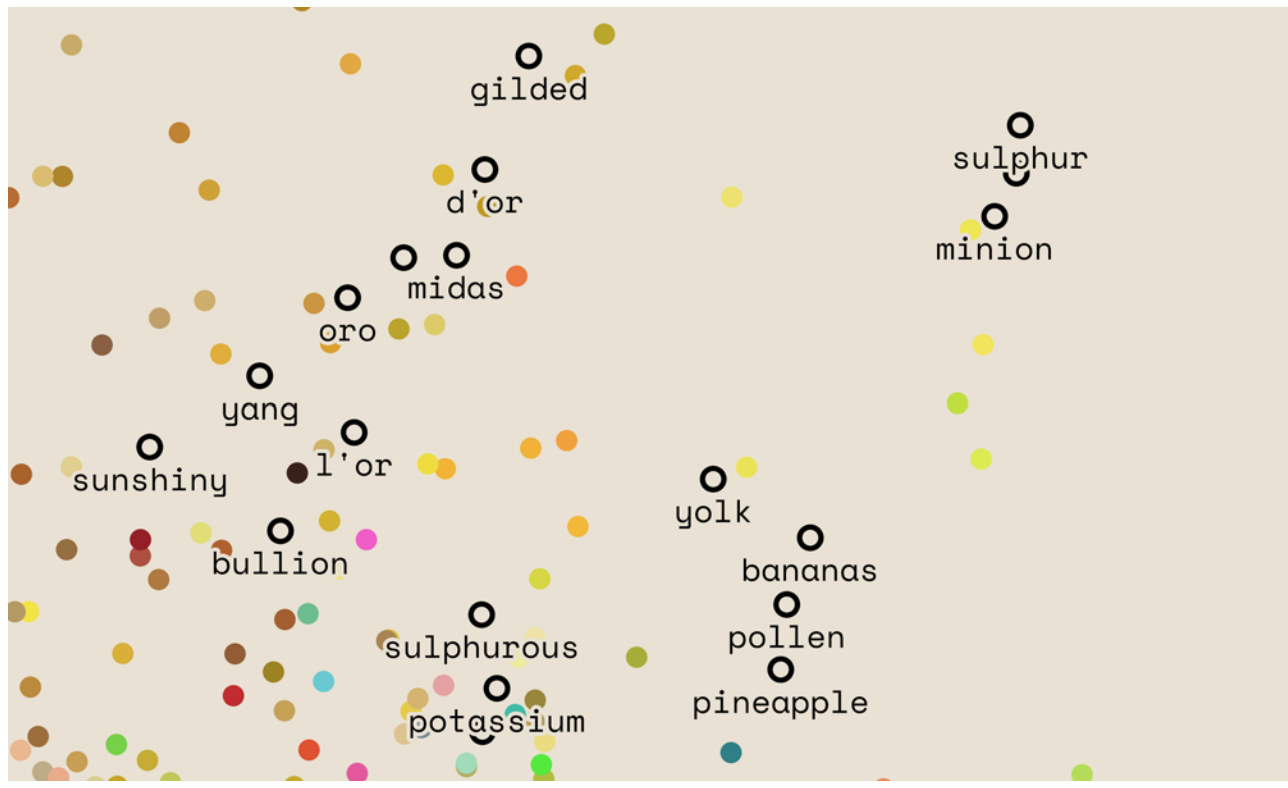

↖ Non-colour terms projected onto the map often land in regions tied to their linguistic associations, hinting at a recoverable colour-adjacent structure in CLIP’s representation.

## DISCUSSION

Colour language offers a useful lens into the semantic and perceptual organisation of large vision-language models. Prior work probing CLIP at the neuron level has shown specific colour deficiencies, including biases against achromatic stimuli and poor performance on colour naming tasks [5]. More broadly, colour categorisation varies across languages in ways that large language models struggle to capture [6].

Rather than evaluate CLIP through task-based prompting, we treat the visualisation itself as a form of explanation: surfacing structure in the model's colour semantics that is otherwise opaque to users. This project began as an artistic exploration of colour naming and developed into a visual study of how CLIP organises colour terms. The visualisation is lightweight and can be ported to SVG and browsers, allowing artists and designers to navigate the resulting 2D map and constellations.

The visualisations expose clusters and transitions that are both legible and meaningful. Some neighbourhoods trace shifts across dusk, sunset, and nightfall. Others connect terms like avocado, tennis ball, and honeydew through shared regions of green. Some nearby terms are closely aligned in perceptual space, while others relate through linguistic, cultural, or metaphorical associations.

For artists and designers, this supports a mode of colour exploration and palette building shaped by naming, mood, material reference, and association.

It offers an alternative to standard colour libraries, tools, and palette books, such as Pantone and HSL colour pickers—which often organise colour primarily by geometry in colour space—and instead allows colour to be browsed as a field of linked semantic and perceptual relations.

These clusters hint at a lower-dimensional manifold within CLIP's colour semantics, and our visualisations demonstrate that this structure could be made legible to designers looking to bring more semantic meaning into their use of colour. This points towards future artist-facing colour tools grounded in foundation models, extending the maps shown here into a more interactive user experience.

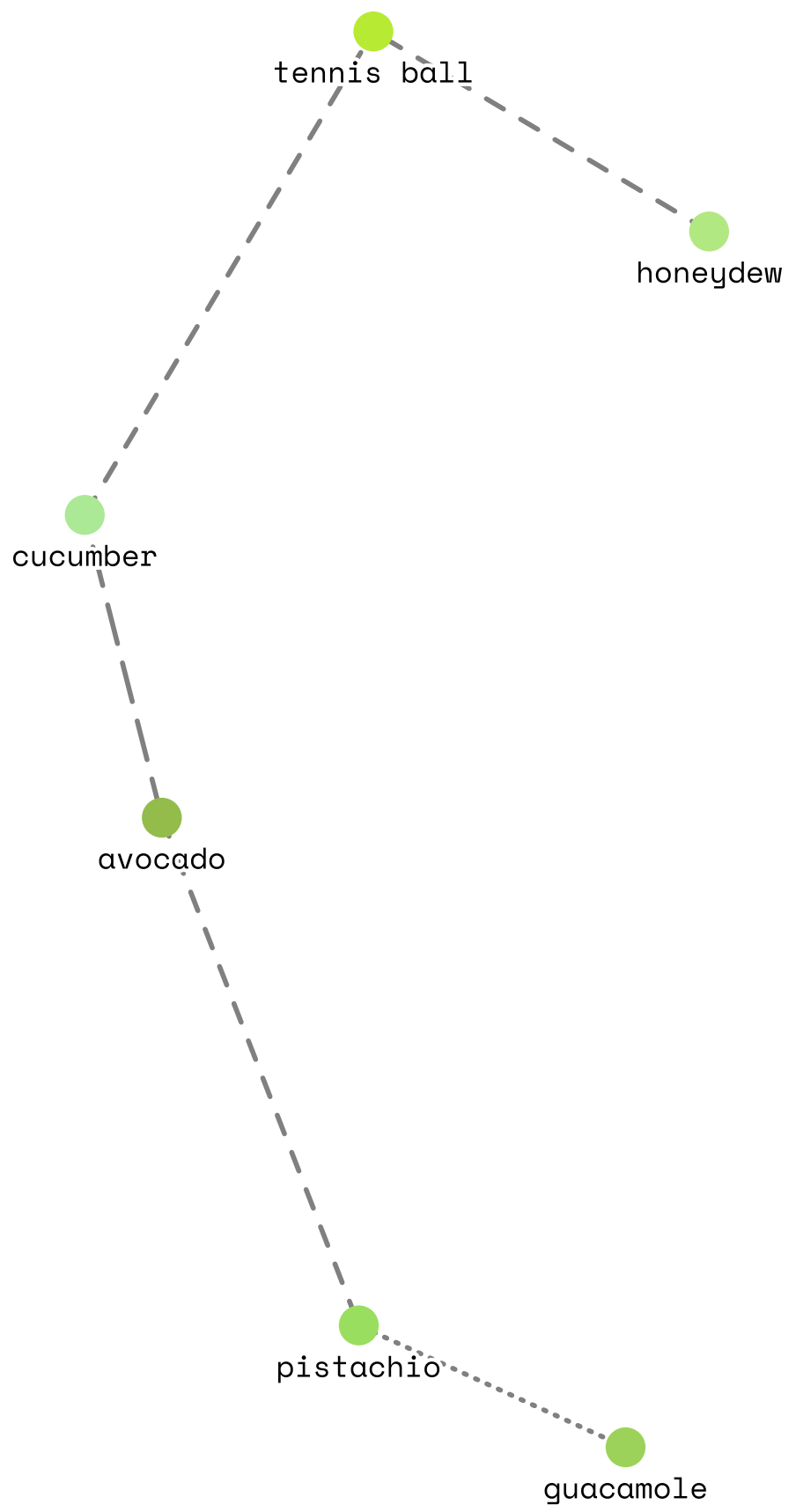

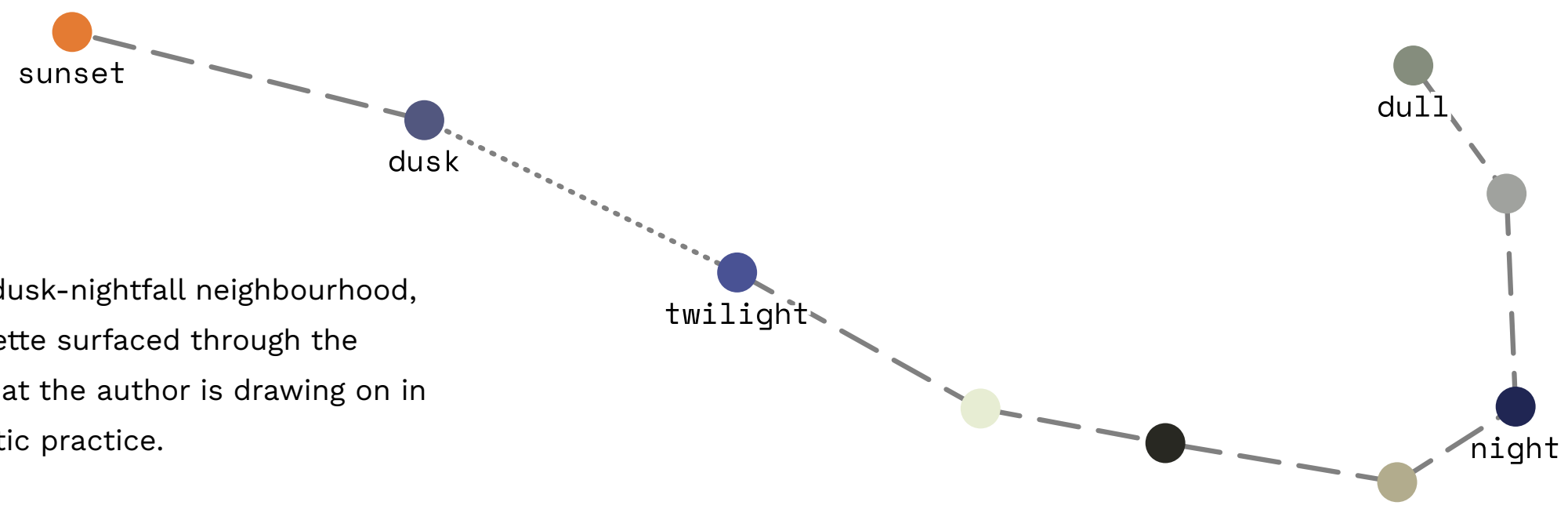

↗ The sunset-dusk-nightfall neighbourhood, a narrative palette surfaced through the visualisation that the author is drawing on in their own artistic practice.